# Magnetic Anisotropy of $SrCu_2(BO_3)_2$ System as Revealed by X-Band ESR

Andrej Zorko,[1,*] Denis Arčon,[1] Hiroshi Kageyama,[2] and Alexandros Lappas[3]

[1]*Institute "Jožef Stefan", Jamova 39, 1000 Ljubljana, Slovenia*

[2] *Department of Chemistry, Graduate School of Science, Kyoto University, Kyoto, 606-8502, Japan*

[3] *Institute of Electronic Structure and Laser, Foundation for Research and Technology – Hellas, Vassilika Vouton, 71110 Heraklion, Greece*

X-band ESR measurements on a single crystal of the highly frustrated $SrCu_2(BO_3)_2$ system are shown to provide an essential inspection of the magnetic anisotropy present in this compound. The very broad absorption lines seem to be consistent with the largest anisotropy term, namely, the antisymmetric Dzyaloshinsky-Moriya (DM) interaction allowed by symmetry. However, the previously well-accepted model of only interdimer interaction is generalized with additional intradimer DM terms. Moreover, spin-phonon coupling is recognized as the cause on the linewidth broadening with increasing temperature.

## 1. Introduction

Low-dimensional spin-gap systems with quantum-disordered singlet ground state have been in the main stream of scientific research in the last decade due the diversity of possible ground state phases and richness of low-lying magnetic excitation spectra. The gap in magnetic excitation spectrum (spin gap) is a common feature in one-dimensional (1D) antiferromagnets and is usually triggered by competing exchange interactions or dimerisation present in the system. On the contrary, the field of spin-gap systems in two dimensions is not nearly as rich as in one dimension since the general characteristic of two-dimensional (2D) antiferromagnets is a tendency to exhibit three-dimensional order at lower temperatures due to interlayer exchange interactions. However, sufficient frustration provided by a special exchange topology of the system can resolve the non-magnetic singlet state as the ground state of the system and is thus directly responsible for the opening of the spin gap.

$SrCu_2(BO_3)_2$ with localized $S=1/2$ $Cu^{2+}$ spins is one of the few 2D systems exhibiting a spin-gap behaviour [1]. Already in the early stages [2] this system was recognised as the first chemical realisation of a model proposed by Shastry and Sutherland, which takes into account 2D square lattice of antiferromagnetically coupled spins with additional selected diagonal bonds [3]. The Hamiltonian of the investigated system is thus given by

$$H_{ex} = J\sum_{(ij)} \mathbf{S}_i \cdot \mathbf{S}_j + J'\sum_{\{lm\}} \mathbf{S}_l \cdot \mathbf{S}_m , \tag{1}$$

where (*ij*) terms represents nearest-neighbour (*nn*) exchange coupling pairs while {*lm*} stands for next-nearest-neighbour (*nnn*) pairs of spins with exchange constants estimated to be $J = 85$ K and $J' = 54$ K [4]. The unique network of orthogonal interacting dimers composed of $Cu^{2+}$ ions is shown as the inset to Fig. 1. The ground state of the system remains a simple product of spin singlets on each dimer even if interlayer exchange coupling $J'' = 8$ K is taken into account [5], which makes the *J'*-coupling extremely frustrated. As a result of the inherent geometrical frustration the lowest lying magnetic excitation – a single triplet excitation present on one of the dimers – is extremely localized as evidenced by almost dispersionless excitation band [6] and relatively small value of the spin gap $\Delta = 34$ K with respect to the exchange coupling constant *J* [6,7].

---

[*] Corresponding author: tel.: +386-1-4773866; fax: +386-1-4263296; e-mail: andrej.zorko@ijs.si .

The singlet-triplet excitation possesses another remarkable property, namely, a fine structure as revelled for the first time by high-field ESR experiment [7]. The interpretation of high-field ESR spectra was, in fact, quite an achievement as far as the understanding of magnetic anisotropy of the system in concerned. The experimental singlet-triplet transitions are forbidden due to the fact that the Shastry-Sutherland exchange Hamiltonian commutes with a magnetic dipole operator. Such transitions can be explained only by including spin-anisotropy terms to the original Hamiltonian, which are then responsible for mixing a small amount of excited states into the ground singlet state. A feasible analysis of the observed angular-dependent fine structure was proposed by Cépas et al. with the introduction of antisymmetric anisotropic exchange interaction, known as Dzyaloshinsky-Moriya (DM) interaction of the form [8]

$$H_{DM} = \sum_{\{ij\}} \pm D_\| \mathbf{e}_c \cdot \mathbf{S}_i \times \mathbf{S}_j . \tag{2}$$

Such antisymmetric interaction is quite common in highly anisotropic, i.e. low-dimensional systems, with large exchange coupling and no centre of inversion between interacting spins. It is the first-order perturbation effect of spin-orbit coupling and can often be by far the largest anisotopic term of the Hamiltonian. The magnitude of the DM interaction can be estimated from the *g*-shift and exchange constant, $D \sim \Delta g/g \cdot J$. As the Cu-B-O planes are almost flat [9] only interdimer DM interaction was employed with Dzyaloshinsky-Moriya vectors $D_\| = 2.1$ K [8] pointing in the crystal **c**-direction (inset to Fig. 1), due to symmetry arguments originally proposed by Moriya [10].

Magnetic anisotropy determines spin dynamics in the system and is thus reflected in ESR spectra through the observed linewidths and lineshifts. To establish the nature of the largest anisotropic terms we previously conducted X-band ESR measurements on powdered sample of $SrCu_2(BO_3)_2$ and concluded that symmetric anisotropic exchange and/or antisymmetric DM interaction had to be considered in order to justify the observed broad spectra [11]. However, performing experiments on single crystals offers an additional insight into the anisotropy on the system. Namely, angular dependence of the linewidth can yield supplementary information on the direction of anisotropy axes in addition to the magnitude of anisotropic contributions. Moreover, the presence of diffusive motion of spins can be verified through the enhanced contributions of secular parts of anisotropic terms. Measurements on single crystals thus often allow one to rule out particular anisotropy terms in favour of other ones. Performing X-band ESR on a single crystal of SrCu$_2$(BO$_3$)$_2$ hence seems a reasonable way to verify the validity of the picture of spin dynamics proposed for this particular system. Furthermore, it can yield information on additional anisotropic terms, which have not yet been evaluated or even observed by other spectroscopic techniques.

## 2. Material and methods

Single crystal of SrCu$_2$(BO$_3$)$_2$ was grown by the travelling solvent floating zone (TSFZ) method from a polycrystalline SrCu$_2$(BO$_3$)$_2$ powder using LiBO$_2$ as a solvent [12]. The direction of crystal axes was determined by means of X-ray scattering and the high purity was verified by almost no low-temperature Curie-like upturn of the magnetic susceptibility down to 2 K. A crystal of a size 5ä2ä2 mm$^3$ was used for ESR experiments.

X-band electron spin resonance (ESR) measurements were performed on a commercially available Bruker E580 FT/CW spectrometer working at Larmor frequency of $\nu_L$ = 9.5 GHz. The temperature of 525 K was reached using high-temperature controller with preheated-nitrogen-flow cryostat.

## 3. Results and discussion

Angular-dependent X-band ESR spectra on a single crystal of SrCu$_2$(BO$_3$)$_2$ were recorded at 295 K and 525 K, i.e. at two temperatures well above the characteristic intradimer exchange temperature $J$ = 85 K. Bearing in mind that the $H_0 = H_z + H_{ex}$ part of the total spin Hamiltonian $H = H_0 + H'$ is dominant, the anisotropic part $H'$ can be treated as a perturbation and the Kubo-Tomita approach to magnetic resonance should be justified at such high temperatures [13]. Here $H_z$ and $H_{ex}$ terms represent Zeeman and isotropic exchange (Eq. (1)) part of Hamiltonian, respectively. As we are dealing with $S$ = 1/2 spins no single-spin anisotropy terms are present. So the magnetic anisotropy is in principal driven by the dipolar interactions ($H_{dd}$) between neighbouring magnetic moments, the hyperfine coupling ($H_{hf}$) of electron spins with nuclear spins, the symmetric anisotropic exchange interaction ($H_{ae}$) (pseudo-dipolar interaction) and the antisymmetric DM interaction ($H_{DM}$), i.e.

$$H' = H_{dd} + H_{hf} + H_{ae} + H_{DM}. \quad (3)$$

The absorption lines should be extremely exchange narrowed as the isotropic exchange is by far the superior interaction present in the investigated system ($k_B J$ à $g\mu_B B_0 \sim g\mu_B \Delta B_{pp}$). In this regime peak-to-peak linewidth is given by the expression [14]

$$\Delta B_{pp} \approx \frac{2}{\sqrt{3}} \frac{k_B^2}{\hbar g \mu_B} \frac{M_2}{\omega_{eff}} \approx \frac{2}{\sqrt{3}} \frac{k_B}{g \mu_B} \frac{M_2}{J}, \quad (4)$$

where $\omega_{eff} = k_B J/\hbar$ is characteristic frequency of exchange modulation and $M_2$ is the second moment, given by

$$M_2 = \langle [H',S_+][S_-,H'] \rangle / \langle S_+ S_- \rangle. \qquad (5)$$

The lineshape is expected to be Lorentzian. However, line broadening and deviations from Lorentzian profiles are often detected in low-dimensional systems at higher temperatures due to spin diffusion [15].

The standard analysis of the lineshape of ESR absorption spectra for $SrCu_2(BO_3)_2$ for both the parallel and the perpendicular direction of external magnetic field **B₀**, with respect to the crystal anisotropy **c**-axis, is shown in Fig. 1 for $T = 525$ K. In the plot $(-(B-B_0)f'_{max}/f'(B)\Delta B_{pp})^{1/2}$ vs. $(2(B-B_0)/\Delta B_{pp})^2$ Lorentzian profiles yield a linear dependence, while Gaussian lines produce much faster (exponentially) increasing function due to rapid decrease of the signal in the tails of resonant absorption spectra. Here $f'(B)$ represents derivative absorption spectrum with peaks $\pm f'_{max}$. Spin diffusion results in a function somewhere in-between the two extreme cases [16]. As it can be seen from Fig. 1 the presence of spin diffusion can be excluded in $SrCu_2(BO_3)_2$ system as the absorption profiles obey Lorentzian dependence.

### 3.1 *g*-factor anisotropy

Next we focus on the anisotropy of the *g*-factor presented in Fig. 2 for $T = 295$ K. As expected from the crystal structure [9] *g*-tensor is axially symmetric with a pronounced angular dependence in the *ac*-plane ($\theta$-dependence) and only minor dependence with respect to the azimuthal angle $\phi$ (*ab*-plane). Fitting the $\theta$-dependence to the equation

$$g = \sqrt{g_\parallel^2 \cos^2\theta + g_\perp^2 \sin^2\theta} \qquad (6)$$

yields parallel and perpendicular *g*-factors with respect to **c**-axis: $g_\parallel = 2.269\pm0.002$ and $g_\perp = 2.057\pm0.002$. Such *g*-shifts are common for $Cu^{2+}$ $3d^9$ configuration where crystal field is stronger than spin-orbit coupling [17, p. 48]. Crystal field thus effectively quenches orbital motion producing an effective spin of 1/2 for the ion. Orbital moments are partially restored by spin-orbit coupling leading to *g*-shifts. The anisotropy of the *g*-shifts in *ab*-plane is much smaller. It can be limited by experimental accuracy to a value smaller than 0.002. At 525 K *g*-factor shows virtually the same angular-dependence with unchanged eigenvalues.

### 3.2 Spin-anisotropy contributions to linewidths

Linewidths of the X-band ESR spectra at 295 K and 525 K are presented in Fig. 3. As already emphasized [11] all the anisotropic terms included in Hamiltonian (3) have to be considered in determination of the contributions to the ESR linewidth. Furthermore, the Lorentzian lineshape of absorption spectra confirms the assumption of the presence of strong exchange narrowing, so that the use of Eq. (4) is well justified. Following the previous estimates of the dipolar and hyperfine contribution to the second moment $M_2^{dd} \approx (g\mu_B/k_B \cdot 1000 \text{ G})^2$ and

$M_2^{hf} \approx (g\mu_B/k_B \cdot 200\text{ G})^2$ [11], linewidths of the order of $\Delta B_{pp}^{dd} \approx 2$ G and $\Delta B_{pp}^{hf} \approx 0.1$ G are expected, which are several orders of magnitude smaller than the experimental values. Employing symmetric anisotropic exchange, which is a result of second order perturbation of spin-orbit coupling (its amplitude can be evaluated as $AE \sim (\Delta g/g)^2 \cdot J = 0.8$ K), and antisymmetric DM interaction yields the second moment of the form [18]

$$M_2^{ae,DM} = \frac{S(S+1)}{3N}\left(2\sum_{(ij)} AE^2 + \sum_{\{lm\}}\left\{(D_{lm}^x)^2 + (D_{lm}^y)^2 + 2(D_{lm}^z)^2\right\}\right), \quad (7)$$

where the components of the DM vector are given in the laboratory frame with the $z$-axis pointing in the direction of the external magnetic field $\mathbf{B_0}$. Using Eq. (4) and Eq. (7) the contribution of the symmetric part can be evaluated as $\Delta B_{pp}^{ae} \approx k_B AE^2/2\sqrt{3}g\mu_B J \approx 20\text{ G}$, while the antisymmetric DM interaction gives a value of $\Delta B_{pp}^{DM} \approx 2k_B D^2/\sqrt{3}g\mu_B J \approx 460\text{ G}$ taking into account the interdimer DM interaction $D_\parallel = 2.1$ K as proposed in ref. 8.

The above analysis favours the DM interaction as the only candidate capable of inducing sufficiently large linewidths, as observed in the current experiment. This conclusion could have been already reached by conducting the experiment on the powdered sample, where the room temperature peak-to-peak linewidth was around 760 G. However, the fundamental advantage of the present single-crystal experiment is the power to test the proposed picture of magnetic anisotropy in SrCu$_2$(BO$_3$)$_2$ system given by Eq. (2) through the angular dependence of the observed linewidths. This can in fact yield additional information about the direction of Dzyaloshinsky-Moriya vectors as discussed below.

### 3.3 Linewidth anisotropy determined by DM interaction

Let the external magnetic field $\mathbf{B_0}$ have an arbitrary orientation described with the polar angle $\theta$ with respect to the crystal $\mathbf{c}$-axis and the azimuthal angle $\phi$ in the $ab$-plane. Applying the transformation of DM vector from laboratory frame to crystal frame gives the expression of the angular dependence of Dzyaloshinsky-Moriya second moment

$$M_2^{DM} = \frac{S(S+1)}{3N}\sum_{\{lm\}}\left\{(D_{lm}^a)^2(1+\sin^2\theta\cos^2\phi) + (D_{lm}^b)^2(1+\sin^2\theta\sin^2\phi) + (D_{lm}^c)^2(1+\cos^2\theta) + \right.$$
$$\left. + 2D_{lm}^a D_{lm}^b \sin^2\theta\sin\phi\cos\phi + 2D_{lm}^a D_{lm}^c \sin\theta\cos\theta\cos\phi + 2D_{lm}^b D_{lm}^c \sin\theta\cos\theta\sin\phi\right\}. \quad (8)$$

Entering only the interdimer DM interaction $\mathbf{D} = (0,0,D_\parallel)$ into Eq. (8) yields the angular dependence of the second moment of the form $M_2^{DM} \propto (1+\cos^2\theta)$. However, the angular dependence of the observed linewidth at 295 K and 525 K is significantly different from this prediction (Fig. 3). $\theta$-dependence can be well described by an equation of the form $A + B(1+\cos^2\theta)$ with $A$ and $B$ contributions of the same order of magnitude. The parameters thus obtained are $A = 478\pm5$ G, $B = 436\pm5$ G for room temperature and $A = 541\pm5$ G, $B = 439\pm5$ G for the higher temperature. The angular-dependent part $B$ is temperature independent and coincides very well with the order-of-magnitude prediction of 460 G, which should be asymptotically approached when going to infinite-temperature. However, unexpectedly, there is an additional term *A*, which is of the same amplitude as the *B* term. On the contrary to the latter, it is also *T*-dependent. The magnetic anisotropy introducing such contribution to the linewidth should be quite strong as the amplitudes of *A* and *B* terms are virtually the same. The symmetric-anisotropy-interaction contribution $\Delta B_{pp}^{ae}$ can be excluded for at least two reasons. The first one is pointed out by the above order-of-magnitude calculation and the second one is due to the characteristic angular dependence of this interaction [19]. Namely, symmetric anisotropic exchange interaction can be presented by a traceless tensor whose principal axes coincide with those of the *g*-tensor. As the SrCu$_2$(BO$_3$)$_2$ system is axially symmetric to a very good approximation, with the anisotropy axis pointing in the crystal **c**-direction, the contribution of $H_{ae}$ to the linewidth is expected to vary with the orientation of external magnetic field according to $M_2^{ae} \propto (1 + \cos^2 \theta)$. This is the same angular dependence as provided by DM interdimer interaction with DM vectors perpendicular to *ab*-plane and as such can not provide an adequate explanation of the additional *A* term observed in the experiment.

We propose that the origin of the experimental angular dependence of the linewidth lies in additional magnetic-anisotropy terms of the Dzyaloshinsky-Moriya type. Although, in the first approximation, treating the crystal *ab*-plane as strictly planar, DM vectors are obliged to point into **c**-direction and no intradimer interaction is allowed due to symmetry reasons, one can loosen these constrains if a buckling of *ab*-planes is taken into account [20]. Bending of neighbouring CuO$_4$ plaquettes forming dimers suppresses inversion symmetry within dimers thus allowing finite intradimer DM interaction $D_\perp$. In addition, also finite in-plane component of interdimer interactions $D_{ab}$ is endorsed. In fact, it has been recently reported that the latter interaction is responsible for the observed anomalies in high-resolution inelastic neutron scattering experiments [21]. A sizable in-plane interdimer component of the order of 30% of $D_\parallel$ was suggested. The ESR linewidth is proportional to the square of the amplitude of DM vector, which would imply that the corrections of the $D_{ab}$ term amount in a correction of only about 10% of the contribution to the linewidth due to $D_\parallel$ interaction. However, the observed effect in our experiment is much larger.

The behaviour of the linewidth in *ab*-plane is less exciting as there is virtually no $\phi$-dependence (Fig. 3). The linewidths seem to be randomly scattered as a function of the azimuthal angle with the amplitude of 5 G around the value of 686 G. Such behaviour corresponds to the addition of intradimer DM interaction since in this case DM vectors are forced to lie in the *ab*-plane in a direction perpendicular to dimers due to symmetry; there is a mirror plane including a dimer and a mirror plane perpendicular to a dimer (see inset to Fig. 1). A detailed analysis of the effect of intradimer DM interaction on the ESR linewidth is currently in progress and will be published elsewhere [22]. Preliminary results suggest that this interaction is about of the same magnitude as the well-excepted interdimer interaction perpendicular to the plane. However, it has to be stressed that in addition to the calculation of the second moment, a detailed analysis of the exchange modulation frequency $\omega_{eff}$ is required since there is a high degree of frustration present in the system. Furthermore, the effect of additional DM interaction on the fine structure of single triplet excitations has to be considered carefully.

At the moment the origin of the intradimer DM exchange interaction seems to some extent unclear. The buckling of *ab*-planes is rather small as the bending angle of $CuO_4$ plaquettes amounts to 8° at room temperature [20]. This angle continuously drops towards 0° approaching the structural phase transition at 395 K so that in the high-temperature phase only $D_\parallel$ interdimer component is again allowed in the static-lattice picture. However, as this interaction alone still does not allow an observation of singlet-triplet transitions in high-field ESR experiments, DM interaction of a dynamical origin was proposed [23]. In this picture lattice-vibrations instantaneously break local symmetry allowing for the Dzyaloshinsky-Moriya anisotropy. It seems that our ESR results would favour the dynamical mechanism, as there is only a small difference in the angular dependence of the linewidth at 295 K and 525 K, i.e. at temperatures well below and well above the structural phase transition, respectively.

The effect of the buckling of *ab*-planes on the anisotropy of g-factor within the planes has also been considered. Since dimers are tilted by half the bending angle with respect to the **c**-axis (4° at room temperature) one can perform a rotation of the *g*-tensor of individual dimer, whose principal axes are tilted due to buckling, into crystal frame. This transformation yields two different components of in-plane *g*-factor when the in-plane external magnetic field is applied, i.e. $g_\perp^a = \left(g_\perp^2 \cos^2\theta + g_\parallel^2 \sin^2\theta\right)^{1/2}$ for the case when magnetic field is parallel to the dimer and $g_\perp^b = g_\perp$ for magnetic field perpendicular to the dimer, where $g_\perp$ and $g_\parallel$ are the *g*-factor components in the eigenframe of *g*-tensor. However, due to rather small corrugation of *ab*-planes the effect is unobservablly small as $g_\perp^a - g_\perp \approx 0.001$. This finding is consistent with almost no $\phi$-dependence of g-factor shown in Fig. 2.

### 3.4 Temperature dependence of linewidths

Finally, let us comment on the temperature dependence of the observed linewidth anisotropy (Fig. 3). The increase of the linewidth at 525 K with respect to room temperature is reverse to the behaviour below room temperature, previously reported for powdered sample [11]. However, the similar increase was also observed in powdered sample when going beyond room temperature. The increase of the linewidth between 295 K and 525 K in $\theta = 0°$ direction is virtually the same as in the case when the external field is applied in $\theta = 90°$ direction. Both directions yield 65±5 G broader lines at 525 K than at 295 K. Trying to give an explanation for such angular-independent behaviour, or the observed line broadening in the first place, one can immediately exclude spin diffusion as the origin of the increase since the lineshape remains "perfectly" Lorentzian even at 525 K. Next step would be the incorporation of static spin correlations, which can play an important role in the temperature evolution of the linewidth in low-dimensional systems and can be sometimes seen in a rather broad temperature range between $T ¥ J$ and $T \sim 10 J$ [14]. In their paper Soos et al. showed that including static spin correlations into an ordinary *nn* square lattice leads to temperature dependent second moment

$$M_2(J,T) = \left(M_2^S F^S(J,T) + M_2^A F^A(J,T)\right) \frac{\chi_C}{\chi(T)}, \tag{9}$$

where indexes $S$ and $A$ stand for spin symmetric and spin antisymmetric contributions, respectively. The infinite-temperature second moments $M_2^{S,A}$ are multiplied by functions describing temperature evolutions of static spin correlations $F^{S,A}(J,T)$ and by the ratio between Curie-law susceptibility $\chi_C$ and the observed one $\chi(T)$. If the increase of the linewidth is due to static spin correlations one will expect to observe angular-dependent increase as the angular dependence is contained in the infinite-temperature second moments and is thus simply multiplied by corrections originating from static spin correlations. The calculation of $M_2(J,T)$ in SrCu$_2$(BO$_3$)$_2$ system is more complicated due to the form of its Hamiltonian and a simple factorisation into high-$T$ second moment and a $T$-dependent function is not possible. However, one still expects to detect angular-dependent increase, which is not the experimentaly observed [22]. Excluding also short-range order effects consequently indicates temperature-independent linewidths with the assumption of pure spin Hamiltonian determining spin dynamics.

Such temperature-independent behaviour is often observed in magnetic crystals with significant exchange narrowing. It is then plausible that spin-phonon coupling has to be taken into consideration in the SrCu$_2$(BO$_3$)$_2$ system in order to account for the increase in the linewidth with the temperature. In fact, normal modes can modulate anisotropic spin interactions via spin-phonon interaction, which then induces transitions between energy levels of the system. The linewidth, regulated by a finite life-time of a spin in a particular energy level, is in this scenario expected to rise with temperature due to the increasing number of phonons. However, the *A* term cannot entirely be of the dynamical origin as one might expect. Its increase between 295 K and 525 K is too small with respect to increase of the convex function describing the number of phonons that characterises the direct phonon process [17, p. 112] of relaxation. Higher-order (Raman) phonon processes are expected to yield even faster temperature-varying broadening of the linewidth with increasing temperature. Therefore we propose that the temperature-dependant part of the *A* term is due to spin-phonon coupling, while its major (constant) part is due to magnetic anisotropy of DM origin. Spin-lattice coupling has indeed been reported to play a crucial role in a dramatic softening of the elastic constants in SrCu$_2$(BO$_3$)$_2$ [24]. It was suggested to be very strong and advised to be taken into consideration in any future model considering this particular system.

## 4. Conclusions

In summary, angular dependence of X-band ESR spectra of SrCu$_2$(BO$_3$)$_2$ a single crystal have been presented at two different temperatures. The magnetic anisotropy governing spin dynamics has been shown to be quite substantial as the observed linewidths are very broad in spite of the presence of strong isotropic exchange coupling. Among the possible anisotropy mechanisms the antisymmetric anisotropic Dzyaloshinsky-Moriya interaction was determined as the only one capable of producing sufficiently large linewidths. However, previously proposed interdimer DM exchange coupling, with DM vectors perpendicular to crystal *ab*-plane, does not suffice for an adequate description of experimentally observed angular dependence. In fact, sizable intradimer DM interaction has to be employed to provide satisfactory explanation. Its origin, though, still remains unclear. It can be due to dynamical effects and/or is a fingerprint of buckling of crystal *ab*-planes. In addition, the angular-independent increase of the linewidth with increasing temperature above room temperature the existence of strong spin-phonon coupling present in the system.

## Acknowledgements

We thank the General Secretariat for Science & Technology (Greece) for the provision of financial support through a Greece-Slovenia 'Joint Research & Technology Program'. AL and HK thank the Yamada Science Foundation (Japan) for supporting their work in low-dimensional materials.

**Figure legends:**

**Fig. 1:** Lineshape analysis for X-band absorption lines of $SrCu_2(BO_3)_3$ single crystal for external magnetic field applied parallel (open circles) and perpendicular (full diamonds) to crystal **c**-axis at 525 K. Solid lines correspond to absorption profiles of Lorentzian and Gaussian type. The inset shows the 2D network of $Cu^{2+}$ ions, with thick lines representing the nearest-neighbor exchange coupling $J$ and thin lines related to next-nearest-neighbor exchange $J'$. The direction of interdimer $D_\parallel$ as well as intradimer $D_\wedge$ DM interaction is also presented.

**Fig. 2:** Anisotropy of the *g*-tensor of the X-band ESR spectra for the $SrCu_2(BO_3)_3$ single crystal in the crystal *ac*-plane ($\theta$-dependence; full circles) and in the *ab*-plane ($\phi$-dependence; open circles) at 295 K. The fit (solid line) corresponds to Eq. 6.

**Fig. 3:** Linewidth anisotropy of the X-band ESR absorption lines for the $SrCu_2(BO_3)_3$ single crystal in the crystal *ac*-plane (full symbols) at 295 K (circles) and 525 K (diamonds) and in the *ab*-plane at 295 K (open circles). Experimental data is fitted to equation of the form $A + B\cos^2\theta$ (solid lines).

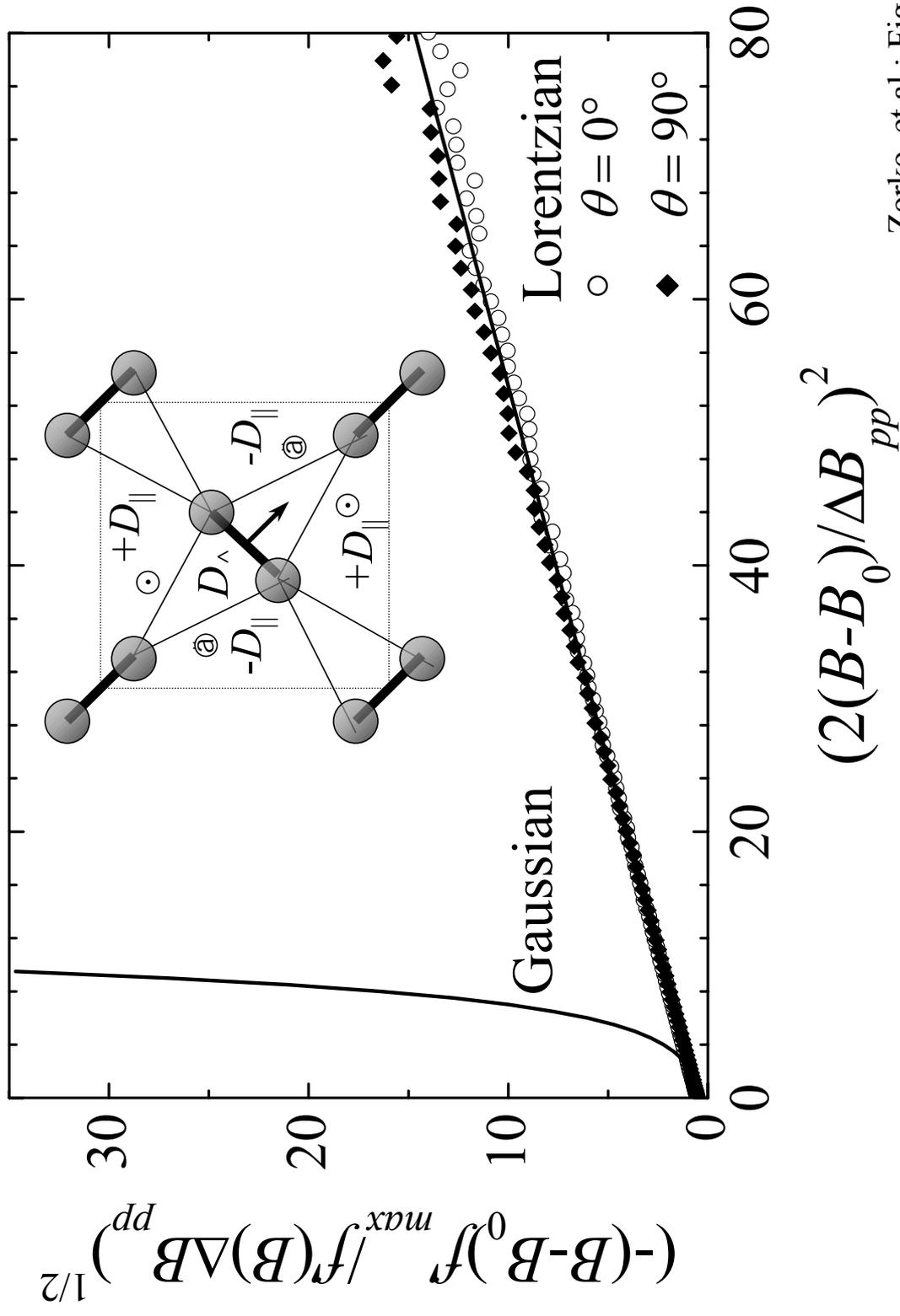

Zorko, et al.: Fig. 1

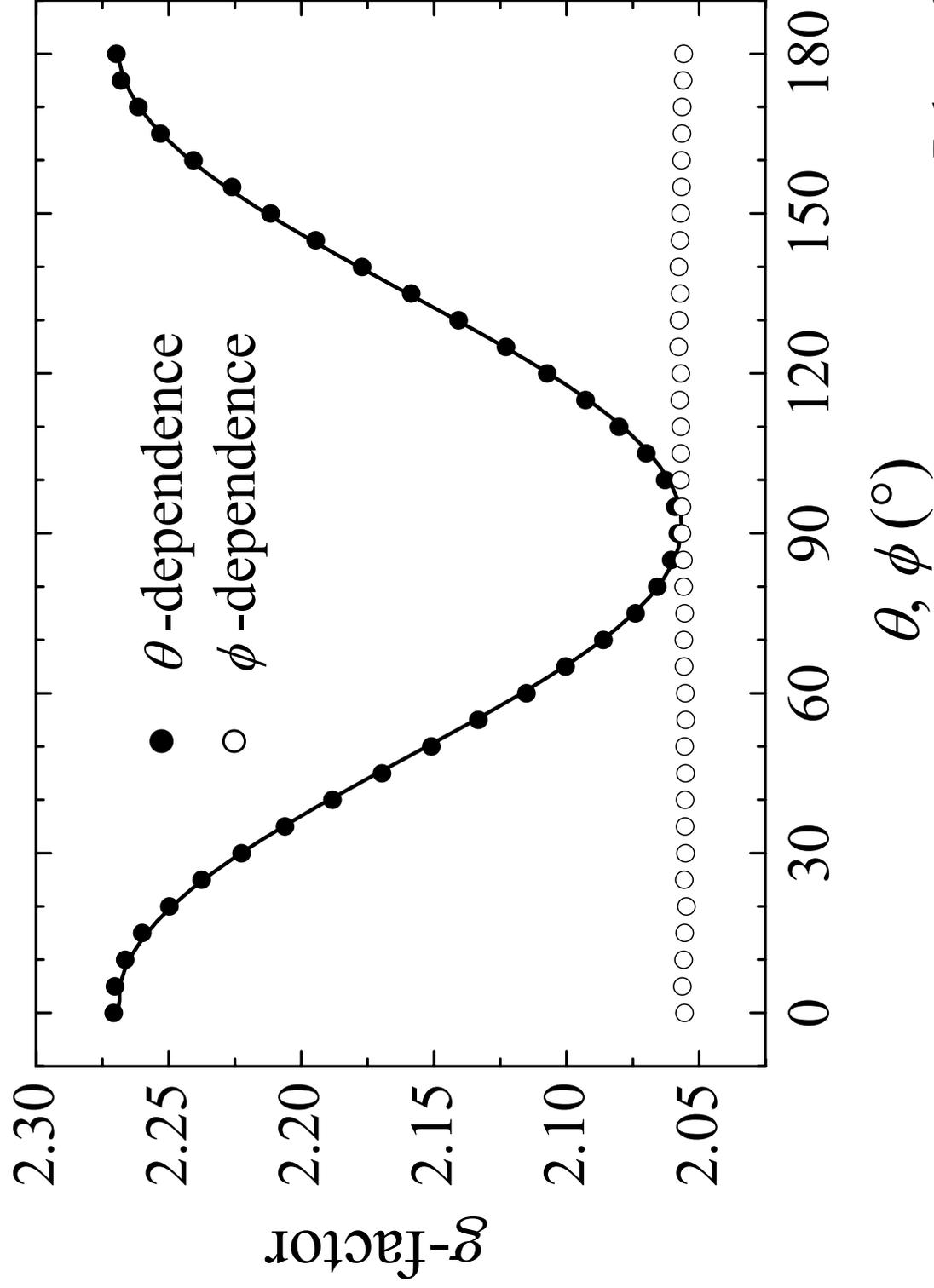

Zorko, et al.: Fig. 2

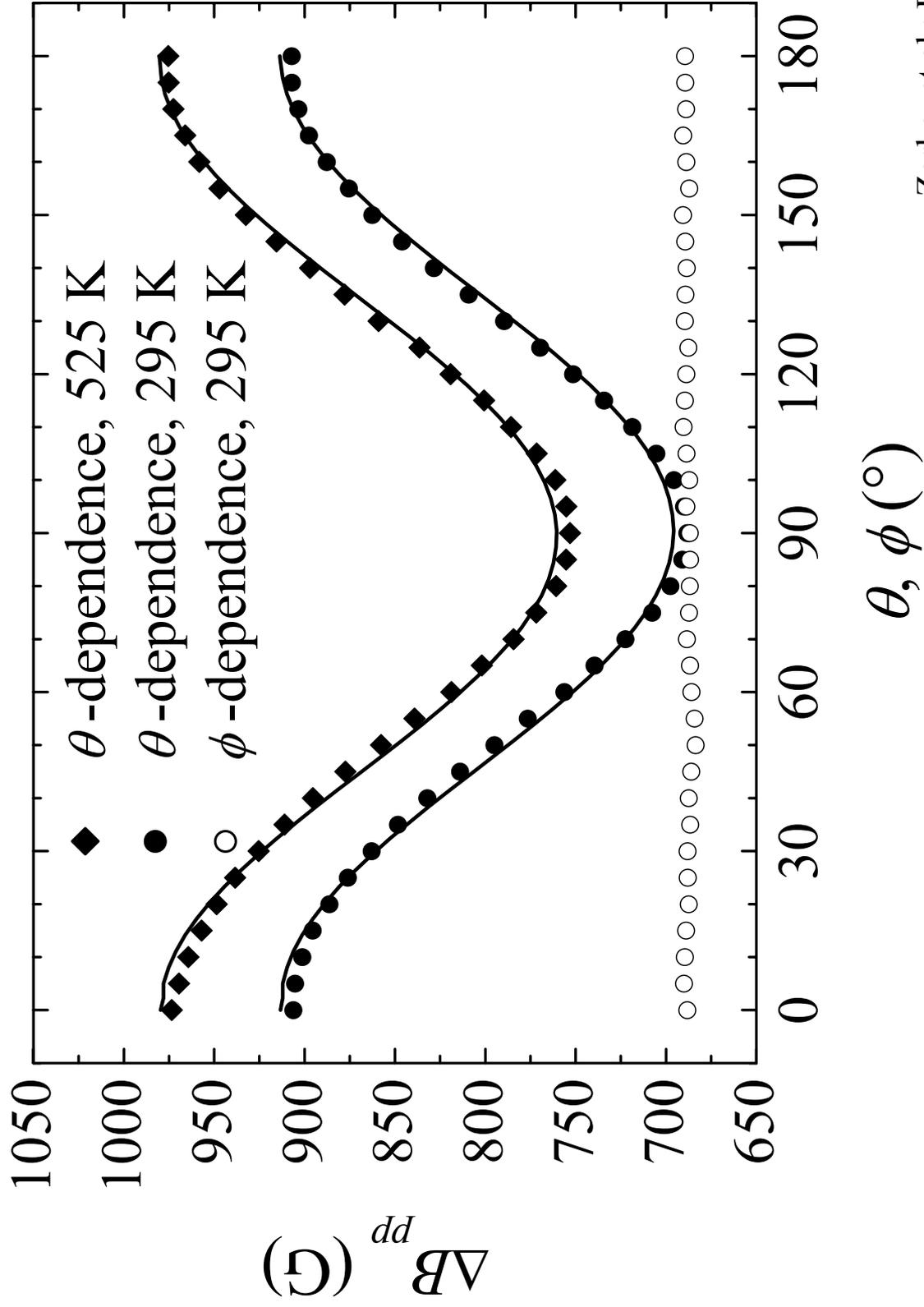

Zorko, et al.: Fig. 3